\documentclass[12pt,a4paper,english,superscriptaddress,aps,preprint]{revtex4}
\usepackage{amsmath}
\usepackage{amssymb}
\usepackage{graphicx}
\usepackage{hhline}
\usepackage{mwe}
\usepackage{amsbsy}
\usepackage{textcomp}

\usepackage{stackengine}
\stackMath

\makeatletter
\usepackage{babel}
\newcommand{\bea}{\begin{eqnarray}}
\newcommand{\eea}{\end{eqnarray}}

\newcommand{\pa}{\partial}

\newcommand{\be}{\begin{equation}}
\newcommand{\ee}{\end{equation}}
\numberwithin{equation}{section}

\begin{document}
\immediate\write16{<<WARNING: LINEDRAW macros work with emTeX-dvivers
                    and other drivers supporting emTeX \special's
                    (dviscr, dvihplj, dvidot, dvips, dviwin, etc.) >>}

\title{\boldmath SUSY Chern-Simons $\mathbb{CP}^N$ and baby Skyrme models and their BPS structures}
\date{30-01-2017}
\author{Jose M. Queiruga}
\email{jose.queiruga@kit.edu}
\affiliation{Institute for Theoretical Physics, Karlsruhe Institute
of Technology (KIT), 76131 Karlsruhe, Germany}
\affiliation{Institut for Nuclear Physics, Karlsruhe Institute of Technology, Hermann-von-Helmholtz-Platz 1,
D-76344 Eggenstein-Leopoldshafen, Germany}

\begin{abstract}
 We construct the $\mathcal{N}=2$ supersymmetric extensions of different models containing a Chern-Simons term and study their BPS structure. We find that the coupling of the Chern-Simons term to the general gauged nonlinear $\sigma$-model allows for general potentials depending on the K\"{a}hler metric and preserving two supersymmetries. We construct from supersymmetry the potentials that allow for BPS structures in different $\mathbb{CP}^{N}$ models coupled to Chern-Simons and Maxwell terms. We also couple the Chern-Simons term to a higher-derivative field theory (the baby Skyrme model) and study its SUSY extension and BPS structure.
  \end{abstract}

\maketitle


\section{Introduction}
\label{intro}

The existence of self-dual vortex solutions in Abelian Chern-Simons systems was studied years ago \cite{Jackiw, Lee1}. The authors found that, for a particular choice of the Higgs potential the vortex satisfies a set of Bogomoln'yi or self-duality equations. It was shown later that this particular potential allowing for self-dual solutions arises naturally once extended supersymmetry (SUSY) is imposed \cite{Lee2,Ivanov}. In addition, in the model with extended SUSY, the central charge of the superalgebra is related to the topological charge (which in turn provides the Bogomoln'yi bound) \cite{Witten1,Hlousek}. These relations between SUSY and self-duality were exploited later in the literature for Chern-Simons systems (see for example \cite{Lee3,Kao}), for the Abelian-Higgs model \cite{Edelstein} or more recently for the baby Skyrme models \cite{Queiruga1, Queiruga2, Queiruga3, Bolognesi} or higher-derivative field theories \cite{Nitta1,Nitta2, Queiruga4, Queiruga5, Queiruga6}.

Once one imposes SUSY, one of the obvious consequences is a reduction of the number of free parameters of the model but also possible geometrical constraints in the target-space manifold \cite{Gaume,Zumino} and even a restriction on the form of the interaction terms. This is the case for the Chern-Simons Higgs systems. When coupled to a linear $\sigma$-model term (i.e. the standard Dirichlet kinetic term), $\mathcal{N}=2$ determines uniquely the potential. We will show here that the coupling of the Chern-Simons term to a general nonlinear $\sigma$-model allows for general potentials even in the $\mathcal{N}=2$ case. This potential will depend on the K{\"a}hler metric, and as a consequence, an infinite family of self-dual potentials (i.e. allowing for self-dual solutions) can be generated depending on the choice of the metric for the nonlinear $\sigma$-model. We will start in the $\mathcal{N}=1$ superspace and obtain the necessary constraints to achieve an $U(1)$ symmetry in the fermionic sector (absence of fermion number violating terms (f.n.v.)). This extra symmetry translates ultimately in an implicit extended SUSY of the model, leading to the desired potentials and self-dual structure. Throughout this paper we will work with $2+1$ dimensional models. We understand therefore that $\mathcal{N}=1$ SUSY means two supersymmetric charges and $\mathcal{N}=2$ means four. 

In addition to the computation of energy bounds and self-dual potentials, once we are equipped with $\mathcal{N}=2$ SUSY, we can systematically determine the BPS (Bogomoln'yi-Prasad-Sommerfeld)  equations without an explicit knowledge of the fermionic sector. We will show how to construct an ``universal" set of BPS equations for gauge theories in three dimensions and provide a method to determine the BPS-type of the solutions.

Then, we apply these ideas to different gauged $\mathbb{CP}^N$ models and compare our results, naturally obtained from SUSY, with previous results found in the literature. The gauged $\mathbb{CP}^N$ models, specially the  $\mathbb{CP}^1$ and  $\mathbb{CP}^2$ models, were treated previously regarding their BPS structure \cite{Zakr1,Izquierdo, NittaV, Loginov, Casana1,Casana2,Casana3}, their SUSY solutions \cite{Zakr2,Zakr3} or some generalized BPS models \cite{Bazeia1, Bazeia2, Bazeia3}. We will construct an $\mathcal{N}=2$ extension for the gauge formulation of the gauge $\mathbb{CP}^N$ models with Chern-Simons term. This formulation provides an explanation, in terms of supersymmetry, to the presence (absence) of self-dual structure in some models.


If higher-derivative terms play a role, the issue of the SUSY extension is far from trivial (see \cite{Khoury1,Khoury2,Nitta1,Nitta2,Farakos1,Farakos2,Nitta3,Nitta4,Queiruga4,Queiruga6,Queiruga66} for SUSY theories with higher-derivative terms), and the general conditions under which a model allows for SUSY extension are still an open question. In the second part of the paper we will focus on a particular higher-derivative field theory, the baby Skyrme model \cite{Piette1,Piette2,Adam1} with a Chern-Simons term. The (ungauged) baby Skyrme model corresponds to the three dimensional restriction of the Skyrme model \cite{Skyrme}. It consists of a quadratic term in derivatives of the $\sigma$-model type, a quartic term (the proper Skyrme term) and a potential. The potential is mandatory to guarantee the existence of solitonic solutions according to the Derrick's theorem. The supersymmetric extensions of the baby Skyrme model have been studied previously \cite{Queiruga1,Queiruga2,Queiruga3,Bolognesi} (for possible SUSY extensions of Skyrme-type models in four dimensions see \cite{Nepomechie, Queirugax, Gud1, Gud2}). When the $\sigma$-model term is absent from the action the resulting model is the so-called BPS (Bogomoln'yi-Prasad-Sommerfeld)  baby Skyrme model and, unlike the original proposal it has a Bogomoln'yi bound and solutions saturating it. This property is also preserved in the gauged version (the Maxwell BPS baby-Skyrme model \cite{Adam2}). We will see here that, in the presence of a Chern-Simons term, an extra term has to be added to the action in order to have BPS solutions. The form of this term is determined by the requirement of $\mathcal{N}=2$ SUSY. It should not be surprising that the sum of two BPS models is not in general BPS. This is only possible when the BPS equations of both models coincide. In the present case, the BPS equations involved (Chern-Simons-$\mathbb{C}P ^1$ and BPS baby Skyrme) do not have common solutions except the trivial one. Therefore, it is natural to expect a modification of the model that brings it back to the BPS limit. We will provide also a classification of the BPS-type of the solutions for various modifications of the model.

This paper is organized as follows: in Sec. 2, we construct a Chern-Simons nonlinear $\sigma$-model in the $\mathcal{N}=1,2$ superspace and analyze the BPS potential. In Sec. 3, we provide a general form of the BPS equations for gauge theories in three dimensions and study the possible BPS-types within our framework. In Sec. 4, we apply the ideas of the previous two sections to various Chern-Simons $\mathbb{CP}^N$ models and compute explicitly the BPS potentials and BPS equations. In Secs. 5 we restrict our analysis to $\mathbb{CP}^1$ and $\mathbb{CP}^2$ Chern-Simons models found in the literature and compare with our results and determine the BPS-type of the BPS solitons. In Secs. 6 and 7 we turn to SUSY gauged higher-derivative field theories (baby Skyrme models) with a Chern-Simons term. By using the superspace formulation we obtain a ``modified" model that is BPS and compute BPS equations and BPS-type for the solutions. Finally, Sec. 7 is devoted to our summary. We also add three appendices with some computational details.


\section{From $\mathcal{N}=1$ to $\mathcal{N}=2$ Chern-Simons with arbitrary potential}
\label{Chern}
The $\mathcal{N}=1$ superspace formulation of the Chern-Simons term can be expressed as follows
\be
\mathcal{L}_{CS}^{\mathcal{N}=1}=-\frac{1}{8\pi}\kappa\int d^2\theta D^\alpha \Gamma^\beta D_\beta \Gamma_\alpha =\frac{1}{4\pi}\kappa \epsilon^{\mu\nu\rho}A_\mu \pa_\nu A_\rho-\frac{1}{4\pi}\kappa \chi^\alpha \chi_\alpha,\label{N1CS}
\ee
where $\kappa$ is the Chern-Simons level and $\Gamma^\alpha$ is a real spinor superfield, which in the Wess-Zumino gauge contains a gauge field $A_\mu$ and a Majorana spinor $\chi_\alpha$. The superderivative $D_\alpha$ in the three dimensional $\mathcal{N}=1$ superspace is defined by
\be
D_\alpha=\frac{\pa}{\pa\theta^\alpha}+i \theta^\beta\sigma^\mu_{\alpha\beta}\pa_\mu,\quad \sigma^\mu_{\alpha\beta}\text{ are the Pauli matrices}.
\ee

As we have mentioned in the introduction, the coupling of this term to the linear $\sigma$-model with two supersymmetries leads irreparably to a fix potential. Therefore, a natural attempt to construct a Chern-Simons action with extended SUSY and a general potential is to consider the general nonlinear $\sigma$-model. In the $\mathcal{N}=1$ superspace form we have
\bea
\mathcal{L}_\sigma^{\mathcal{N}=1} &=&-\frac{1}{2}\int d^2\theta g\left(\Phi^\star\Phi\right)\left(D^\alpha+ie\Gamma^\alpha\right)\Phi^\star \left(D_\alpha-ie \Gamma_\alpha\right)\Phi=\nonumber \\
&&g(\phi^\star\phi)\left(\vert D_\mu\phi\vert^2+i \psi^{\alpha\star}\sigma^{\mu\,\beta}_{\alpha}\pa_\mu\psi_\beta+i e \left(\psi^{\alpha\star}\lambda_\alpha\phi-\lambda^\alpha \psi_\alpha\phi\right)+F^\star F\right)+\nonumber\\
&&+\frac{1}{2}g''(\phi^\star\phi)\vert\phi \vert^2\left(\psi^{\alpha\star}\psi_\alpha \right)^2+\frac{1}{2}g'(\phi^\star\phi)\left(F^\star\phi+F\phi^\star+2\psi^{\alpha\star}\psi_\alpha\right)\psi^{\alpha\star}\psi_\alpha-\nonumber\\
&&-\frac{1}{2}g'(\phi^\star\phi)\left(\psi^{\beta\star}\phi+\psi^\beta\phi^\star\right)\left(\left(i\pa_\beta^{\,\alpha}\phi^\star+C^\alpha_{\,\beta}F^\star\right)\psi_\alpha-\psi^{\alpha\star}\left(i\pa_{\beta\alpha}\phi+C_{\alpha\beta}F\right)\right)-\nonumber\\
&&-i\frac{ e}{2}g'(\phi^\star\phi) \left(\psi^{\beta\star}\phi+\psi^\beta\phi^\star\right)\sigma^{\mu\,\alpha}_{\beta}A_\mu\left(\phi^\star\psi_\alpha-\psi^\star_\alpha\phi\right).\label{nsm}
\eea
where $\Phi^\star,\Phi$ are complex superfields constructed in terms of $\mathcal{N}=1$ real superfields
\bea
\Phi&=&\sigma_1+i\sigma_2,\,\, \Phi^\star=\sigma_1-i\sigma_2\\
\sigma_1&=&\text{Re}\phi+ \theta^\alpha\text{Re}\,\psi_\alpha-\theta^2\text{Re}\,F,\\
\sigma_2&=&\text{Im}\phi+ \theta^\alpha\text{Im}\,\psi_\alpha-\theta^2\text{Im}\,F.
\eea
The next step is to analyze the fermionic sector of the resulting model. The invariance under the $\mathcal{N}=2$ SUSY requires the symmetry $\psi\rightarrow e^{i\alpha}\psi$ for all complex spinors. As a consequence, a new real spinor $\rho^\alpha$, has to be introduced in such a way that it can be combined with $\chi^\alpha$ to form a complex spinor
 \be
\Sigma^\alpha=\rho^\alpha+i \chi^\alpha, \quad \Sigma^{c\, \alpha}=\rho^\alpha-i \chi^\alpha.
\ee
An straightforward analysis of the Lagrangian (\ref{N1CS}) suggests a quadratic term of the form 
\be
\mathcal{L}_{a1}=-\frac{\kappa}{4\pi}\int d^2\theta S^2=-\frac{\kappa}{2\pi}\sigma D-\frac{\kappa}{4\pi}\rho^\alpha\rho_\alpha,\label{extra1}
\ee
where $S$ is a real superfield $S=S(\sigma,\rho^\alpha,D)$. It is easy to see that the last term in (\ref{N1CS}) combines with the last term in (\ref{extra1}) to form a complex spinor. It is also necessary to compensate the real spinor $\lambda^\alpha$ in (\ref{nsm}) with an appropriate term containing $\rho_\alpha$. We propose the following term 
\bea
\mathcal{L}_{a2}&=&\lambda\int d^2\theta S h (\Phi^\star\Phi)=\lambda Dh (\phi^\star\phi)+\lambda\sigma h'(\phi^\star\phi)\left(F\phi^\star+ F^\star\phi+2\psi^{\alpha\star}\psi_\alpha\right)+\nonumber\\
&&+\lambda\frac{1}{2}\sigma h''(\phi^\star\phi)\left(\psi^{\alpha\star}\psi_{\alpha\star}\phi^2+\psi^\alpha\psi_\alpha\phi^{\star 2}+2\psi^{\alpha\star}\psi_\alpha\phi^\star\phi\right)+\nonumber\\
&&+\lambda h'(\phi^\star\phi)\rho^\alpha\left(\psi^\star_\alpha\phi+\psi_\alpha\phi^\star\right).
\eea

In terms of the complex fermion $\Sigma$, the f.n.v. terms of the full Lagrangian $\mathcal{L}^{\mathcal{N}=1}$ ($\mathcal{L}=\mathcal{L}_{CS}^{\mathcal{N}=1}+\mathcal{L}^{\mathcal{N}=1}_{\sigma}+\mathcal{L}_{a1}+\mathcal{L}_{a2})$ can be cast as
\be
\mathcal{L}^{\mathcal{N}=1}\vert_{\text{f.n.v.}}=\frac{1}{2}\left(\lambda h'-e g\right)\left(\phi^\star\Sigma^\alpha\psi_\alpha+\text{h.c.} \right)+\left(\phi^2\frac{\pi\lambda}{\kappa g}\left(g h''-g' h'\right)\psi^{\alpha\,\star}\psi^\star_\alpha+\text{h.c.}\right).
\ee

It is clear that the symmetry $\psi\rightarrow e^{i\alpha}\psi,\,\, \Sigma\rightarrow e^{i\alpha}\Sigma$ (and therefore the absence of f.n.v. terms) is achieved for
\be
g=C h' \,\, \text{and} \,\, \lambda=C e,
\ee
where $C$ is a constant. As usual, extended SUSY constrains the number of coupling constants and, in addition, sets the function $h$ to be a derivative of the $\sigma$-model metric. After eliminating the auxiliary fields, the bosonic part of the action can be written as follows 
\be
\mathcal{L}^{\mathcal{N}=1}\vert_{\psi=\Sigma=0}=\frac{\kappa}{4\pi} \epsilon^{\mu\nu\rho}A_\mu \pa_\nu A_\rho+g\vert D_\mu\phi\vert^2-\frac{4\pi^2 e^4}{\kappa^2}g h^2\vert\phi\vert^2\label{cs12},
\ee
where the potential depends now on the K\"{a}hler metric coming from the $\sigma$-model term. We recover the standard result for $g=1$ \cite{Lee2} (the linear $\sigma$-model) with the characteristic $\vert\phi\vert^6$ potential. We can also straightforwardly introduce a non-zero vacuum expectation value for $\phi$ by means of a Fayet-Iliopoulos term. In general, and with $\mathcal{N}=2$, the choice of the K\"{a}hler metric determines uniquely the potential and vice versa. We have learnt also that, in order to have extra SUSY in three dimensional gauge models in the $\mathcal{N}=1$ superspace, the real spinor superfield $\Gamma_\alpha$ has to be accompanied by a real superfield $S$ in a specific combination. Of course, if we content ourselves with $\mathcal{N}=1$ (i.e. two supercharges), no extra superfields are necessary and the Chern-Simons Lagrangian (\ref{N1CS}) does not need to be ``compensated" and, as a consequence, we have complete freedom in the choice of the potential.

 The formulation in the $\mathcal{N}=2$ superspace is much simpler.  We simply have to couple the $\mathcal{N}=2$ Chern-Simons term to the general gauged nonlinear $\sigma$-model
\be
\mathcal{L}_\sigma^{\mathcal{N}=2}=\int d^4\theta \mathcal{K}\left(\Phi^\dagger e^V \Phi\right)\label{sigmaact},
\ee
where $\Phi$ ($\Phi^\dagger$) are chiral (antichiral) superfields and $V$ is the vector superfield
\be
V=\theta\sigma^\mu\bar{\theta}A_\mu+i\theta\theta\bar{\theta}\lambda-i\bar{\theta}\bar{\theta}\theta\bar{\lambda}-\frac{1}{2}\theta\theta\bar{\theta}\bar{\theta}D+\theta \gamma^5\bar{\theta}\sigma\label{vector}.
\ee
The last term in (\ref{vector}) comes from the dimensional reduction from four to three dimensions and corresponds to the third component of the gauge field. This field $\sigma$ corresponds to the lowest component of the real superfield $S$ in the $\mathcal{N}=1$ formulation. The $\mathcal{N}=2$ Chern-Simons Lagrangian has the following form \cite{Ivanov}
\be
\mathcal{L}_{CS}^{\mathcal{N}=2}=-i\sum_{\alpha=\dot{\alpha}}\int d^4\theta \frac{\kappa}{2\pi}\bar{D}^{\dot{\alpha}}V D_\alpha V\label{N2CS},
\ee
where
\be
\mathcal{D}_\alpha=D_\alpha+e D_\alpha V,\quad \bar{\mathcal{D}}_{\dot{\alpha}}=\bar{D}_{\dot{\alpha}}+e\bar{D}_{\dot{\alpha}} V.
\ee

Taking into account that 
\be
\sum_{\alpha=\dot{\alpha}} \lbrace D_\alpha, \bar{D}^{\dot{\alpha}}  \rbrace=0,
\ee
the model possesses the usual gauge symmetry
\bea
\Phi &\rightarrow& e^{-i t\Lambda}\Phi, \quad \bar{D}_{\dot{\alpha}}\Lambda=0,\\
\Phi^\dagger &\rightarrow & e^{i t\Lambda^\dagger}\Phi^\dagger, \quad D_\alpha\Lambda^\dagger=0,\\
V &\rightarrow & V+ i \left(\Lambda-\Lambda^\dagger\right).
\eea

After substituting the auxiliary fields in $\mathcal{L}^{\mathcal{N}=2}=\mathcal{L}_\sigma^{\mathcal{N}=2}+\mathcal{L}_{CS}^{\mathcal{N}=2}$ we just obtain (\ref{cs12}), but now the potential can be written directly in terms of the K\"{a}hler potential from (\ref{sigmaact}),
\be
V=\frac{ e^4\pi^2}{\kappa^2}\vert\phi\vert^2 \pa^2_{\phi,\bar{\phi}}\mathcal{K}\left(\phi\pa_\phi\mathcal{K}+\text{h.c.}\right)^2\label{potCS}.
\ee

When possible, the $\mathcal{N}=2$ superspace formulation is in general more geometrical, and as we will see, it has a close connection with the BPS structure. On the other hand, in situations where the model is know not to posses extended symmetry, the $\mathcal{N}=1$ superspace in the only way to provide a minimal amount of supersymmetry. We will use both formulation of the Chern-Simons action though the following sections.


\section{General BPS equations}
\label{secBPS}

As we have already mentioned, the BPS structure is three dimensions is closely related to the $\mathcal{N}=2$ SUSY. Before we discuss specific models we will see that it is possible to obtain, due to the supersymmetric structure, a set of model-independent or ``universal" BPS equations and we will develop simple criteria to determine the BPS-type.

In a gauge model in three dimensions with $\mathcal{N}=2$ SUSY, the supersymmetric transformation of the complex fermions in the chiral supermultiplet can be written in the following form
\be
\delta\psi=\left(-i\sigma^\mu D_\mu \phi-e\sigma \phi\right)\bar{\xi}+F\xi\label{susytrans}.
\ee
Equivalently, in matrix notation, we can define
\be
\mathcal{M}=
\left(\begin{matrix}
F& 0 & -e\phi\left(\sigma+A_0\right)&i D_{\bar{z}} \phi\\
0& F&iD_z\phi &-e\phi\left(-\sigma+A_0\right)\\
e\bar{\phi}\left(\sigma-A_0\right)&-i D_z \overline{\phi}&\overline{F}&0\\
-iD_{\bar{z}}\overline{ \phi}&e\bar{\phi}\left(\sigma-A_0\right)&0&\overline{F}
\end{matrix}\right), \label{M}
\ee
where $D_z\phi=\left(D_1+iD_2\right)\phi,\,D_{\bar{z}}\phi=\left(D_1-iD_2\right)\phi$. The transformations (\ref{susytrans}) are then
\be
\left(\begin{matrix}
\delta \psi \\
\delta\bar{\psi}
\end{matrix}\right)
=
\mathcal{M}
\left(\begin{matrix}
 \xi \\
\bar{\xi}
\end{matrix}\right).
\ee

For the complex fermion in the gauge supermultiplet the SUSY transformations are
\be
\delta\lambda=-\frac{1}{2}\epsilon^{\mu\nu\rho}F_{\mu\nu}\sigma_\rho\epsilon-D\epsilon+i\sigma^\mu\pa_\mu\sigma\epsilon.
\ee

Again we define the matrix of the transformation
\be
\mathcal{N}=
\left(\begin{matrix}
-i(D-F_{12})& i(\pa_1 A_0+\pa_1\sigma)+(\pa_2A_0+\pa_2\sigma) \\
 i(\pa_1 A_0-\pa_1\sigma)-(\pa_2A_0-\pa_2\sigma) & -i(D+F_{12})
\end{matrix}\right),\label{N}
\ee
therefore
\be
\left(\begin{matrix}
\delta \lambda \\
\delta\bar{\lambda}
\end{matrix}\right)
=
\left(\begin{matrix}
\mathcal{N}&0\\
0&\overline{\mathcal{N}}
\end{matrix}\right)
\left(\begin{matrix}
 \xi \\
\bar{\xi}
\end{matrix}\right).
\ee

An interesting feature of the configurations saturating a Bogomoln'yi bound (or verifying the BPS equations) is that only a fraction, rather than all,  of the supersymmetric generators are broken. This condition corresponds to the vanishing determinant of the matrix transformations (\ref{M}) and (\ref{N})
\bea
\det\mathcal{M}&=&0\Leftrightarrow F\bar{F}=\frac{1}{2}\left(D_z\bar{\phi}D_{\bar{z}}\phi+D_z\phi D_{\bar{z}}\bar{\phi}\right)+e^2\phi\bar{\phi}(A_0^2-\sigma^2)\label{gen11}\\
&\pm&\sqrt{-\mathcal{T}^2-4e^2(\sigma^2-A_0^2)j_zj_{\bar{z}}},\nonumber\\
\det\mathcal{N}&=&0\Leftrightarrow \left(D^2-F_{12}^2\right)^2-2\left(D^2-F_{12}^2\right)\left((\pa_1 A_0)^2-(\pa_1\sigma)^2+(\pa_2A_0)^2-(\pa_2\sigma)^2\right)^2\\
&&-\left((\pa_1A_0-\pa_1\sigma)^2+(\pa_2A_0-\pa_2\sigma)^2\right)\left((\pa_1A_0-\pa_1\sigma)^2+(\pa_2A_0-\pa_2\sigma)^2\right)=0\nonumber, \label{gen22}
\eea
where $\mathcal{T}= D_1\bar{\phi} D_2 \phi-D_1 \phi D_2\bar{\phi}$ and $j_z=\bar{\phi}D_z\phi-\phi D_z\bar{\phi}$. As we announced, these equations are universal in the sense that they only depend on the supersymmetric transformation of the fermions and not on the particular model. If we impose $\sigma=A_0$ (this equation is always satisfied, see appendix C) we arrive at
\bea
F\bar{F}&=&\left(D_1\phi\mp i D_2\phi\right)\left(D_1\bar{\phi}\pm i D_2\bar{\phi}\right),\label{gen1}\\
D&=&\pm F_{12},\label{gen2}\\
\sigma&=&\pm A_0.\label{sigma0}
\eea

In this formulation, the BPS equations (\ref{gen1})-(\ref{sigma0}) contain the information about the model in the particular form of the auxiliary fields. That is to say, given an $\mathcal{N}=2$ gauged model, the knowledge of the explicit form of the auxiliary fields is sufficient to determine the BPS equations.
Within this frame work one can also obtain the amount of SUSY preserved by the BPS soliton, i.e. the BPS-type. Given a BPS soliton, we say that it is $f/q$-BPS or of  $f/q$-type if it preserves $f$ supercharges out of $q$ ($q$ is the total number of supercharges). At this point we stress the fact that a solution of (\ref{gen1})-(\ref{sigma0}) is a BPS soliton and all BPS solitons are solutions of (\ref{gen1})-(\ref{sigma0}), no matter the amount of SUSY they preserve. To determine the BPS-type we proceed as follows. Let $\mathcal{M}_0$ and $\mathcal{N}_0$ be the on-shell matrix transformations (\ref{M}) and (\ref{N}) on a solution of (\ref{gen1})-(\ref{sigma0}). We have the following relation
\be
f=\min\left(\dim \text{ker}\mathcal{M}_0,\dim \text{ker}\mathcal{N}_0\right).\label{frac}
\ee
Phrased differently, for a solution of the BPS equations the dimension of the null space of the matrix transformations gives the number of free Grassmann parameters of the SUSY transformation verifying $\delta\psi_{\text{on-shell}}=\delta\lambda_{\text{on-shell}}=0$ and therefore the number of unbroken supercharges. Moreover, the on-shell structure of (\ref{N}) and the BPS equation (\ref{gen2}) imply that $\dim \text{ker}\mathcal{N}_0\geq 2$. On the other hand, from the on-shell transformation (\ref{M}) with the BPS equations (\ref{gen1}) and (\ref{sigma0}) we get $\dim \text{ker}\mathcal{M}_0\leq 2$. Therefore, the relation (\ref{frac}) reduces to 
\be
f=\dim \text{ker}\mathcal{M}_0.
\ee
It still remains to proof that $\langle  \text{ker}\mathcal{M}_0 \rangle \subseteq \langle  \text{ker}\mathcal{N}_0 \rangle$. This can be done constructively by considering all possibilities. For two-dimensional nontrivial configurations ($F_{12}\neq0$ and $\phi\neq \text{constant}$) there are only two possibilities according to the value of the auxiliary field $F$. For $F=0$ we have from (\ref{gen1})-(\ref{sigma0}) (upper sign):
\bea
\langle \text{Ker}\mathcal{N}_0\rangle&=&\langle\bar{\xi}^{\dot{2}},\xi_1 \rangle,\quad D=F_{12}\neq0,\,A_0=\sigma,\\
\langle\text{Ker}\mathcal{M}_0\rangle&=&\langle\bar{\xi}^{\dot{2}},\xi_1 \rangle,\quad F=\bar{F}=0,\, D_{\bar{z}}\phi=0,\,D_z\bar{\phi}=0,
\eea 
and for the lower sign
\bea
\langle\text{Ker}\mathcal{N}_0\rangle&=&\langle\bar{\xi}^{\dot{1}},\xi_2 \rangle,\quad D=-F_{12}\neq0,\,A_0=-\sigma,\\
\langle\text{Ker}\mathcal{M}_0\rangle&=&\langle\bar{\xi}^{\dot{1}},\xi_2 \rangle,\quad F=\bar{F}=0,\, D_z\phi=0,\,D_{\bar{z}}\bar{\phi}=0,
\eea 
and therefore $\langle  \text{ker}\mathcal{M}_0 \rangle = \langle  \text{ker}\mathcal{N}_0 \rangle$. For $F\neq 0$ we have
\bea
\langle \text{Ker}\mathcal{N}_0\rangle&=&\langle\bar{\xi}^{\dot{2}},\xi_1 \rangle,\quad D=F_{12}\neq0,\,A_0=\sigma,\\
\langle\text{Ker}\mathcal{M}_0\rangle&=&\langle\bar{\xi}^{\dot{2}}=-ie^{-i\,\theta}\xi_1 \rangle,\quad F=e^{i\, \theta}D_{\bar{z}}\phi,\,\bar{F}=e^{-i\, \theta}D_z\bar{\phi},
\eea 
and for the lower sign
\bea
\langle\text{Ker}\mathcal{N}_0\rangle&=&\langle\bar{\xi}^{\dot{1}},\xi_2 \rangle,\quad D=-F_{12}\neq0,\,A_0=-\sigma,\\
\langle\text{Ker}\mathcal{M}_0\rangle&=&\langle\bar{\xi}^{\dot{1}}=-ie^{-i\,\theta}\xi_2 \rangle,\quad F=e^{i\, \theta}D_z\phi,\,\bar{F}=e^{-i\, \theta}D_{\bar{z}}\bar{\phi},
\eea 
and therefore $\langle  \text{ker}\mathcal{M}_0 \rangle \subset \langle  \text{ker}\mathcal{N}_0 \rangle$. From this simple analysis we deduce that the BPS type in this dimension depends basically on the matrix $\mathcal{M}_0$ and on the value of the auxiliary field $F$ (see Table \ref{BPStype}). It is important to note that we are considering two-dimensional static solutions, i.e. $\phi=\phi(x_1,x_2)$. For domain-wall configurations with $\phi=\phi(x_1)$ we still can have $1/2$-BPS states for $F\neq 0$ \cite{Nitta1}.

\begin{table}
\begin{center}
  \begin{tabular}{ | c | c | c | c | }
    \hline
   Auxiliary fields & $\dim \text{Ker}\mathcal{N}_0$ & $\dim \text{Ker}\mathcal{M}_0$&BPS-type \\ \hline
   $D=0,\, F=0,\,\sigma=0$ &4&2&$1/2$\\ \hline
      $D=0,\, F=0,\,\sigma\neq0$ &2&2&$1/2$\\ \hline
         $D\neq0,\, F=0,\, \sigma_{=}^{\neq}0 $ &2&2&$1/2$\\ \hline
              $D=0,\, F\neq0,\,\sigma=0$ &4&1&$1/4$\\ \hline
                $D\neq0,\, F\neq0,\, \sigma_{=}^{\neq}0$ &2&1&$1/4$\\ \hline

  \end{tabular}
  \caption{Classification of the BPS-type according to the auxiliary field values.}
    \label{BPStype}  
\end{center}

\end{table}

\section{The Chern-Simons $\mathbb{CP}^{N}$ model: geometric and gauged formulation.}
\label{geo}

In this section we will focus on a particular type of nonlinear $\sigma$-model, the $\mathbb{CP}^{N}$ model. The geometric formulation of the pure $\mathbb{CP}^{N}$ model is well-known. In the framework of the nonlinear $\sigma$-models the only ingredient needed is the target space manifold metric which we take as the Fubini-Study metric
\be
g_{i\bar{j}}=\frac{\left(1+\vert \boldsymbol{\phi}\vert^2\right)\delta_{i\bar{j}}-\bar{\phi}_i\phi_j}{\left(1+\vert \boldsymbol{\phi}\vert^2\right)^2},
\ee
where $\boldsymbol{\phi}=\left(\phi_1,...,\phi_N\right)$ is an $N$ component complex vector. The (bosonic) $\mathbb{CP}^{N}$ model can be immediately expressed as
\be
\mathcal{L}=g_{i\bar{j}}\pa_\mu\phi_i \pa^\mu\bar{\phi}_{\bar{j}}.
\ee

The $\mathcal{N}=2$ SUSY form is also well-known and can be written as a D-term just containing the K\"{a}hler  potential 
\be
\mathcal{L}_{\mathbb{CP}^{N}}^{\mathcal{N}=2}=\int d^2\theta d^2\bar{\theta}\mathcal{K}(\Phi_i,\Phi^\dagger_{\bar{j}}), \quad \mathcal{K}(\Phi_i,\Phi^\dagger_{\bar{j}})=\ln\left(1+\Phi_i \Phi^\dagger_{\bar{i}}\right).
\ee

The gauged version of the model can be constructed as usual: the K\"{a}hler  potential is replaced by its gauge invariant generalization \cite{Witten2}
\be
\ln\left(1+\Phi_i \Phi^\dagger_{\bar{i}}\right)\rightarrow\ln\left(1+\Phi_i e^{-eV}\Phi^\dagger_{\bar{i}}\right),
\ee
and the corresponding forms of the Yang-Mills and/or Chern-Simons terms are added in an ${\mathcal{N}=2}$ invariant way. For simplicity we will work with $U(1)$ gauge fields from now on. This formulation is very natural and allows us to use all the standard machinery of the SUSY nonlinear $\sigma$-models. There is, however, and alternative formulation of the $\mathbb{CP}^{N}$ model, the so-called gauged formulation. This formulation simplifies sometimes the analysis of various models (see Sec. \ref{examples}).  In constructing the Lagrangian we start with $N+1$ complex fields  $\boldsymbol{\phi}=\left(\phi_1,...,\phi_{N+1}\right)$ and the constraint 
\be
\sum_{i=1}^{N+1}\vert \phi_i\vert^2=1.\label{unit}
\ee
 We have thus $2N+2$ degrees of freedom (d.o.f.) and one constraint, leading to $2N+1$ d.o.f., but from the previous formulation of the model we know that it contains $2N$ d.o.f.. The elimination of the extra d.o.f. is achieved by an $U(1)$ gauging
\be
\mathcal{L}_{\mathbb{CP}^{N}}=D_\mu \phi_i D^\mu\bar{\phi}_i,
\ee
where $D_\mu=\pa_\mu+i A_\mu$. Since $A_\mu$ enters algebraically in the Lagrangian it can be eliminated by using its e.o.m.
\be
A_\mu=\frac{i}{2}\bar{\phi}_i \overleftrightarrow{\pa_\mu}\phi_i.
\ee
We obtain finally
\be
\mathcal{L}_{\mathbb{CP}^{N}}=\nabla_\mu \phi_i \nabla^\mu\bar{\phi}_i,\quad \nabla_\mu=\pa_\mu-\frac{1}{2}\bar{\phi}_i \overleftrightarrow{\pa_\mu}\phi_i.
\ee

The SUSY version of the gauged formulation of the model proceeds as usual, but now, we have to promote the constraint $\sum_{i=1}^{N+1}\vert \phi_i\vert^2=1$ to a SUSY invariant one, resulting in a set of constraints for the vector and chiral superfields (see for example \cite{Nepomechie}). The superfield form in the gauged formulation can be written as
\be
\mathcal{L}_{g. \mathbb{CP}^{N}}^{\mathcal{N}=2}=\int d^2\theta d^2\bar{\theta}\left(-V+\Phi_i^\dagger e^{-V}\Phi_i\right).\label{cpngeo}
\ee

We remark that this model has a global $SU(N)$ symmetry and a local $U(1)$ symmetry which removes the extra d.o.f.. More subtle is what happens in the gauge version of the gauged formulation of the model, i.e. after the addition of a Chern-Simons or Maxwell terms. Now, we require that the model has a $U(1)\times U(1)$ local gauge symmetry (in addition to the previous global symmetry). One of the local gauge symmetries still reduces one d.o.f. while the other is the ``usual" gauge symmetry. In order to achieve the extra local symmetry one extra vector superfield is mandatory in the superspace formulation leading to
\be
\mathcal{L}=\int d^2\theta d^2\bar{\theta}\left(-2e\zeta V-v+\Phi_i^\dagger e^{-eQV-v}\Phi_i\right)+\mathcal{L}_{CS}+\mathcal{L}_{YM},\label{cpngauge2}
\ee
where the first term in (\ref{cpngauge2}) corresponds to the Fayet-Iliopoulos term, $\mathcal{L}_{CS}$ and $\mathcal{L}_{YM}$ are the $\mathcal{N}=2$ Chern-Simons and Maxwell terms respectively and $Q$ is a diagonal charge matrix which will be defined later. Due to the extra $U(1)$ symmetry, the components of the new vector superfield $v=v(a_\mu,\lambda,d)$ are constraint by the SUSY generalization of  (\ref{unit}). After integration in (\ref{cpngauge2}) the auxiliary field $d$ is absent from the action and the only relevant constraint to the (bosonic) action (in addition to (\ref{unit})) is
\be
a_\mu=-\frac{i}{2}\phi_i\overleftrightarrow{D_\mu}\phi_i+\text{ferm.}
\ee

The bosonic part of the Lagrangian (\ref{cpngauge2}) reads in components
\be
\mathcal{L}=-e\zeta D+\nabla_\mu \bar{\boldsymbol{\phi}}\nabla^\mu \boldsymbol{\phi}+D\bar{\boldsymbol{\phi}} Q\boldsymbol{\phi}+\bar{\boldsymbol{F}}\boldsymbol{F}+e^2 \sigma^2\left(\bar{\boldsymbol{\phi}}Q^2\boldsymbol{\phi}-\left(\bar{\boldsymbol{\phi}}Q\boldsymbol{\phi}\right)^2\right)+\mathcal{L}_{CS}+\mathcal{L}_{YM},\label{doublegauge}
\ee
where
\bea
\boldsymbol{\phi}&=&\left(\phi_1,...,\phi_{N+1}\right),\\
\nabla_\mu \boldsymbol{\phi}&=& D_\mu\boldsymbol{\phi}-(\bar{\boldsymbol{\phi}} D_\mu\boldsymbol{\phi})\boldsymbol{\phi},\\
D_\mu\boldsymbol{\phi}&=&\pa_\mu\boldsymbol{\phi}-ie A_\mu Q\boldsymbol{\phi}.
\eea

As pointed out in Sec. 3, the BPS structure of the model can be read from the auxiliary field equations simply by replacing in (\ref{gen1}) $D_i$ by $\nabla_i$,
\bea
\bar{F}_i F_i &=&\nabla_i \phi_a \nabla_i \bar{\phi}_a \pm i\tilde{Q}_a, \quad \tilde{Q}_a=\nabla_1\bar{\phi}_a\nabla_2 \phi_a-\nabla_2\bar{\phi}_a\nabla_1 \phi_a,\label{gen11}\\
F_{12}&=&\mp D.\label{gen12}
\eea

The e.o.m for $F_i$ reads $F_i=0$, regardless the presence of $\mathcal{L}_{CS}$ or $\mathcal{L}_{YM}$. From (\ref{gen11}) we obtain
\be
\left(\nabla_1\pm i\nabla_2\right)\phi_a=0,\label{genBPS11}
\ee
which is the first BPS equation. The second BPS equation (\ref{gen12}) depends of the gauge part of the Lagrangian. We distinguish between three cases: Chern-Simons  $\mathbb{CP}^{N}$, Maxwell $\mathbb{CP}^{N}$ and Chern-Simons-Maxwell $\mathbb{CP}^{N}$. In Table \ref{tabpotentials} we present our results for potentials and gauge BPS equations (via \ref{gen12}) derived from SUSY.  

\begin{table}
\begin{center}
  \begin{tabular}{ | c | c | c | }
    \hline
   $\mathcal{N}=2$ models  & $D$ & $V(\bar{\boldsymbol{\phi}},\boldsymbol{\phi})$ \\ \hline
    CS $\mathbb{CP}^{N}$ & $\frac{4e^3\pi^2}{\kappa^2}\left(\zeta-\bar{\boldsymbol{\phi}}Q\boldsymbol{\phi}\right)\left(-\bar{\boldsymbol{\phi}}Q^2\boldsymbol{\phi}+\left(\bar{\boldsymbol{\phi}}Q\boldsymbol{\phi}\right)^2\right)$ &  $\frac{4e^4\pi^2}{\kappa^2}\left(\zeta-\bar{\boldsymbol{\phi}}Q\boldsymbol{\phi}\right)^2 \left(\bar{\boldsymbol{\phi}}Q^2\boldsymbol{\phi}-\left(\bar{\boldsymbol{\phi}}Q\boldsymbol{\phi}\right)^2\right)$ \\ \hline
    M $\mathbb{CP}^{N}$ & $e\left(\zeta-\bar{\boldsymbol{\phi}}Q\boldsymbol{\phi}\right)$ & $\frac{e^2}{2}\left(\zeta-\bar{\boldsymbol{\phi}}Q\boldsymbol{\phi}\right)^2$ \\ \hline
    MCS $\mathbb{CP}^{N}$ & $\frac{1}{2}\left(\zeta-\bar{\boldsymbol{\phi}}Q\boldsymbol{\phi}\right)-\frac{\kappa}{2\pi}\sigma$ & \parbox[t]{5cm}{$-\frac{1}{8\pi^2}\left(2\pi e\left(\bar{\boldsymbol{\phi}}Q\boldsymbol{\phi}-\zeta\right)+\kappa\sigma\right)^2$\\$+e^2\sigma^2\left(\bar{\boldsymbol{\phi}}Q^2\boldsymbol{\phi}-\left(\bar{\boldsymbol{\phi}}Q\boldsymbol{\phi}\right)^2\right)$}\\ \hline
  \end{tabular}
  \caption{Auxiliary fields and potentials obtained form the $\mathcal{N}=2$ SUSY for three models: Chern-Simons $\mathbb{CP}^{N}$ (CS $\mathbb{CP}^{N}$), Maxwell $\mathbb{CP}^{N}$ (M $\mathbb{CP}^{N}$)  and Maxwell-Chern-Simons $\mathbb{CP}^{N}$ (MCS $\mathbb{CP}^{N}$). The nontrivial BPS equations can be read from the second column, taking into account that $F_{12}=\mp D$.}
    \label{tabpotentials}  
\end{center}

\end{table}

We point out that the presence of the neutral field $\sigma$ is a natural consequence of the extended SUSY. In the Chern-Simons  $\mathbb{CP}^{N}$, $\sigma$ plays de role of an auxiliary field and can be eliminated while in the other two cases $\sigma$ is a genuine dynamical  field. 

We emphasize that the potentials in Table \ref{tabpotentials} are the only ones for which the models present a BPS structure, and of course, they are dictated by the $\mathcal{N}=2$ SUSY simply by solving the e.o.m.'s for the auxiliary fields. In this sense, SUSY provides an explanation for the specific forms of the potentials found in the literature for which there exist or not BPS equations \cite{Loginov,Casana1,Casana2,Casana3,Zakr1,Bazeia1,Bazeia2,Bazeia3, Kimm1,Kimm2}.
  \section{examples}
  \label{examples}

In this section we present a number of applications of the results found in Secs. \ref{secBPS} and \ref{geo}. It often happens that the existence of BPS equations for a given model is intimately linked to a concrete choice of potential. When one considers a bosonic model (without SUSY) the task of looking for this potential can be cumbersome. As we have seen in Sec. \ref{geo} the specific form of the potential as well as the BPS equations are merely a consequence of the extended SUSY.  In the following subsections we apply these results to concrete models found in the literature.

\subsection{$O(3)$ or $\mathbb{CP}^1$  gauged model}

The $O(3)$ Chern-Simons model has the following Lagrangian \cite{Kimm1, Kimm2}
\be
\mathcal{L}=\frac{	\tilde{\kappa}}{2}\epsilon^{\mu\nu\rho}A_\mu\pa_\nu A_\rho+\frac{1}{2}\left(D_\mu\boldsymbol\phi\right)^2-\frac{1}{2\tilde{\kappa}^2}\left(\zeta-\boldsymbol n\cdot\boldsymbol\phi  \right)^2\left(\boldsymbol n \times \boldsymbol\phi\right)^2,\label{Schr}
\ee
where
\be
D_\mu \boldsymbol\phi=\pa_\mu\boldsymbol\phi+A_\mu \boldsymbol n \times \boldsymbol\phi,\quad \boldsymbol n =\left(0,0,1\right),
\ee
and $\boldsymbol \phi$ is a three-component unit vector field. With the specific choice of the potential the model possesses the following BPS equations
\bea
D_1 \boldsymbol \phi&=&\mp \boldsymbol \phi \times D_2\boldsymbol \phi,\label{BPSsch1}\\
F_{12}&=&\pm \frac{1}{\tilde{\kappa}^2}\left(\zeta-\boldsymbol n\cdot\boldsymbol\phi  \right)\left(\boldsymbol n \times \boldsymbol\phi\right)^2.\label{BPSsch2}
\eea

This model is equivalent to the Chern-Simons $\mathbb{CP}^1$ model and both can be connected after the identification $\boldsymbol\phi=\bar{\boldsymbol z}  \boldsymbol\sigma \boldsymbol z$, where $\boldsymbol \sigma$ are the Pauli matrices and here $\boldsymbol z$ corresponds to the fields in the gauged $\mathbb{CP}^1$ model. The potential in (\ref{Schr}) is the potential found in Table 1 for the Chern-Simons $\mathbb{CP}^1$ with $Q=\sigma^3$ and $N=1$. The first BPS equation (\ref{BPSsch1}) is equivalent to (\ref{genBPS11}) while (\ref{BPSsch2}) can be read from Table 1 and (\ref{gen12}) with the replacements $\tilde{\kappa}=\kappa/(2\pi)$ and $e=1$. Since $F=0$, the solutions of (\ref{BPSsch1})-(\ref{BPSsch2}) are $1/2$-BPS. We can consider also the gauged  $O(3)$ model with a Maxwell term whose Lagrangian is \cite{Schroers}
\be
\mathcal{L}=-\frac{1}{4}F_{\mu\nu}F^{\mu\nu}+\frac{1}{2}\left(D_\mu\boldsymbol\phi\right)^2-\frac{1}{2}\left(1-\boldsymbol n\cdot \boldsymbol \phi\right)^2.\label{Schr1}
\ee

The first BPS equation is again (\ref{BPSsch1}) and the second can be read from the third row in Table \ref{tabpotentials}. They are in agreement with \cite{Schroers} for $\zeta=1$ and $e=1$. In this case, we also obtain from \ref{BPStype} that the BPS solutions are $1/2$-type.

\subsection{$\mathbb{CP}^2$ models}
It is well-known that the $\mathbb{CP}^2$ models have topological solitons \cite{Loginov,Casana1,Casana2,Casana3}. In some particular cases, i.e. for a specific choice of the potential, these models possess a self-dual structure. Let us consider the following gauge $\mathbb{CP}^2$ model 
\be
\mathcal{L}=-\frac{1}{4}F_{\mu\nu}F^{\mu\nu}+\left(P_{ab}D_\mu \phi_b\right)^\star P_{ac}D^\mu\phi_c-V(\boldsymbol\phi),\label{cas1}
\ee
where $P_{ab}=\delta_{ab}-\phi_a\phi_b^\star$. According to Table \ref{tabpotentials}, the potential that allows for BPS solutions has the form:
\be
V(\boldsymbol\phi)=\frac{e^2}{2}\left(\zeta-\bar{\boldsymbol\phi}Q\boldsymbol\phi\right)^2.
\ee

As we have mentioned, the first BPS equation is not modified by the gauge part of the action, therefore for the Maxwell $\mathbb{CP}^2$ model still has the form (\ref{genBPS11}). The second BPS equation can be read directly from Table \ref{tabpotentials}. In order to compare with previous results we choose $\zeta=-\frac{1}{4}$ and $Q=\frac{1}{2}\text{diag} (1,-1,0) $. We have
\be
F_{12}=\mp \frac{e}{4}\left(1+2\bar{\phi}_1\phi_1-2\bar{\phi}_2\phi_2\right)
\ee
which is in agreement with \cite{Casana1}. We can consider the Lagrangian as in (\ref{cas1}) but replacing the Maxwell term by a Chern-Simons term. A new inspection of Table \ref{tabpotentials} gives us the potential and the BPS equations. The new one has the generic form
\be
F_{12}=\mp\frac{4e^3\pi^2}{\kappa^2}\left(\zeta-\bar{\boldsymbol{\phi}}Q\boldsymbol{\phi}\right)\left(-\bar{\boldsymbol{\phi}}Q^2\boldsymbol{\phi}+\left(\bar{\boldsymbol{\phi}}Q\boldsymbol{\phi}\right)^2\right).
\ee
Again the specific choice $\zeta=-\frac{1}{4}$ and $Q=\frac{1}{2}\text{diag} (1,-1,0)$ reproduces the results of \cite{Casana2}. 

We can consider finally the Maxwell-Chern-Simons $\mathbb{CP}^2$ model \cite{Casana3}. As already mentioned, in order to have a BPS structure the introduction of a neutral field $\sigma$ is mandatory. In terms of the $\mathcal{N}=2$ SUSY the presence of this field is understood as coming from the dimensional reduction of the gauge field from four to three dimensions. The kinetic term for $\sigma$ appears from the dimensionally reduced Maxwell term and the coupling to the Chen-Simons term eliminates the symmetric solutions $\sigma=0$ (unless $\bar{\boldsymbol\phi}Q\boldsymbol\phi=\zeta$). As a consequence, $\sigma$ appears in the potential (see Table \ref{tabpotentials})
\be
V(\boldsymbol\phi)= -\frac{1}{8\pi^2}\left(2\pi e\left(\bar{\boldsymbol{\phi}}Q\boldsymbol{\phi}-\zeta\right)+\kappa\sigma\right)^2+e^2\sigma^2\left(\bar{\boldsymbol{\phi}}Q^2\boldsymbol{\phi}-\left(\bar{\boldsymbol{\phi}}Q\boldsymbol{\phi}\right)^2\right),
\ee
and also in the BPS equation
\be
F_{12}=\mp\frac{1}{2}\left(\zeta-\bar{\boldsymbol{\phi}}Q\boldsymbol{\phi}\right)-\frac{\kappa}{2\pi}\sigma.
\ee
As in the previous cases, the first BPS equation is not modified (because $F^a=0$) and the solutions are of $1/2$-type. This result can be immediately generalized to all gauged nonlinear K\"{a}hler $\sigma$-models coupled to Chern-Simons or Maxwell actions.


\subsection{A $\mathbb{CP}^1$ type model}
We will finally consider a different type of model in order to give a negative result about the existence of the BPS structure. The $\mathbb{CP}^1$ type model considered in this section does not have a self-dual structure and we will show that it cannot be extended to $\mathcal{N}=2$ SUSY. The non-existence of such an extension is, at the same time, the origin of the non-existence of a self-dual structure. The Lagrangian in which we are interested is
\be
\mathcal{L}=\frac{\kappa}{4\pi}\epsilon^{\mu\nu\rho}A_\mu F_{\nu\rho}+D_\mu\bar{\boldsymbol\phi}D^\mu\boldsymbol\phi-V(\bar{\boldsymbol\phi},\boldsymbol\phi),\label{zak}
\ee
where $\boldsymbol\phi$ is a complex two-component field, $\bar{\boldsymbol\phi}\cdot\boldsymbol\phi=1$ and $D_\mu\boldsymbol\phi=\pa_\mu-iA_\mu\boldsymbol\phi$. This definition of the gauge derivative reduces the local gauge symmetry from $U(1)\times U(1)$ (model (\ref{doublegauge})) to $U(1)$. Note that also is different from (\ref{Schr}) since $\boldsymbol\phi$ is a complex two-component field. It is known that this model does not have self-dual solutions \cite{Zakr1}. The possible $\mathcal{N}=2$ extension of the model is straightforward 
\be
\mathcal{L}=\int d^2\theta d^2\bar{\theta}\left(- V+ \Phi_i^\dagger e^{-V}\Phi_i\right)+\mathcal{L}_{CS}.
\ee

Note that is of the form (\ref{cpngeo}) plus a Chern-Simons term. The difference with respect to (\ref{cpngauge2}) is clear: the local gauge symmetry has been reduced, as explained in the previous paragraph, and the extra gauge d.o.f (entering as an extra vector superfield in (\ref{cpngauge2})) is lost. On the other hand, SUSY invariance of the constraint $\bar{\boldsymbol\phi}\cdot\boldsymbol\phi=1$ requires
\be
A_\mu=\frac{i}{2}\bar{\phi}_i \overleftrightarrow{\pa_\mu}\phi_i+\text{ferm.},
\ee
but this condition is not consistent with the e.o.m. for the gauge field (from (\ref{zak}))
\be
A_\mu=\frac{i}{2}\bar{\phi}_i \overleftrightarrow{\pa_\mu}\phi_i-\frac{\kappa}{4\pi} \epsilon_{\mu\nu\rho}F^{\nu\rho},
\ee
unless $F_{\mu\nu}=0$. We conclude that an $\mathcal{N}=2$ extension of (\ref{zak}) does not exist (except maybe for trivial configurations), justifying at the same time, the non-existence of self-dual structure. The replacement in (\ref{zak}) of the Chern-Simons term by a Maxwell term leads to the same conclusion.
However, it still exists a formulation in the $\mathcal{N}=1$ superspace. The Chern-Simons part is given by (\ref{N1CS}) and the quadratic term in derivatives is a nonlinear $\sigma$-model with target space $\mathbb{S}^3$ (which in the $\mathcal{N}=1$ superspace takes basically the form of (\ref{nsm})). The potential term is simply introduced by a prepotential of the form $\int d^2\theta V(\phi,\bar{\phi})$. This shows, in particular, that $\mathcal{N}=1$ is not enough to guarantee the existence the existence of BPS solitons.


\section{$N=2$ Chern-Simons baby-Skyrme model}
\label{SUSY}

Our goal in this section is to explore the possible BPS structure of the Chern-Simons baby-Skyrme model
\be
\mathcal{L}=g(u\bar{u})D_\mu \bar{u}D^\mu u-h(u\bar{u})\left((D_\mu \bar{u}D^\mu u)^2-\vert D_\mu u D^\mu u \vert^2\right)+\frac{\kappa}{4}\epsilon^{\mu\nu\rho} A_\mu\pa_\nu A_\rho-V(u,\bar{u}) . \label{bSCS1}
\ee
where $u$ is a complex field, $g(u\bar{u})$ is the target space metric of $\mathbb{C}P^1$ and $h(u\bar{u})$ is the area form defined by
\be
g(u\bar{u})=\frac{1}{(1+u\bar{u})^2},\quad h(u\bar{u})=\frac{1}{(1+u\bar{u})^4}.
\ee
On the one hand, we know that the combination of the quadratic, quartic and potential term (the baby-Skyrme model) does not posses BPS equations, and it is hard to believe that the gauging with a Chern-Simons term can change that. On the other hand, for a specific choice of the potential, the combination of the quadratic ($\mathbb{CP}^1$ term) and the Chern-Simons term has BPS equations (for the complex field $u$ they are the gauge generalization of the Cauchy-Riemann equations). But in this case, the possible BPS equations of the remaining quartic and potential terms are not of the Cauchy-Riemann form. It seems that, if the Chern-Simons baby-Skyrme has a BPS structure, something has to be added to (\ref{bSCS1}). We will solve this problem is this section with the help of SUSY. 

The $\mathcal{N}=1$ extension of the Chern-Simons term does not depend on the auxiliary fields (\ref{N1CS}) (for more details see for example \cite{Lee2}). Therefore, the coupling of this term to any other model only changes the bosonic sector precisely by adding the Chern-Simons term. This allows us to construct an $\mathcal{N}=1$ version of the Chern-Simons baby Skyrme model with arbitrary potential (see \cite{Queiruga1,Queiruga2} for the construction of the $\mathcal{N}=1$ baby Skyrme model). Unfortunately, there are no solutions of this model verifying BPS equations. In terms of SUSY, this could be explained by the fact that the model does not allow for an $\mathcal{N}=2$ extension. Hence, the obvious step to explore the possible BPS estructure of the model is the reformulation of the full model in the $\mathcal{N}=2$ superspace. The Skyrme model part consists of two terms, namely the $\sigma$-model term and a quartic term in derivatives 
\bea
\mathcal{L}_\sigma&=&\int d^4\theta \mathcal{K}\left(\Phi^\dagger e^V \Phi\right)\label{sigma},\\
\mathcal{L}_{4}&=&\frac{1}{16}\int d^4\theta h\left(\Phi^\dagger e^V \Phi\right)\mathcal{D}^\alpha\Phi\mathcal{D}_\alpha\Phi\bar{\mathcal{D}}^{\dot{\alpha}}\Phi^\dagger\bar{\mathcal{D}}_{\dot{\alpha}}\Phi^\dagger\label{skBPS}.
\eea
where $\mathcal{D}^\alpha$ and $\bar{\mathcal{D}}^{\dot{\alpha}}$ are the gauge covariant superderivatives. In the superfield formulation, the Lagrangian contains a term quadratic in derivatives (the $\sigma$-model), but in the on-shell Lagrangian this part dissapear leading to the BPS baby Skyrme model. Once one adds the $\mathcal{N}=2$ invariant Chern-Simons term (\ref{N2CS})  the full Lagrangian takes the form
\be
\mathcal{L}=\mathcal{L}_\sigma+\mathcal{L}_4+\mathcal{L}_{CS}\label{bSCS2}.
\ee
Or in components
\bea
\mathcal{L}_\sigma&=&g(u\bar{u})\left(D^\mu u D_\mu \bar{u}+F\bar{F}\right)+e\Omega D-g(u\bar{u})e^2\sigma^2 u\bar{u}+\text{(fermions)},\\
\mathcal{L}_{bS}&=& h(u\bar{u}) \left(\left(D_\mu u \right)^2 \left( D_\nu \bar{u}\right)^2+2 F\bar{F} D_\mu uD^\mu\bar{u} +(F\bar{F})^2 \right.\nonumber\\ 
&&+\left.e^4 \sigma^4 (u\bar{u})^2-2 F\bar{F} e^2 \sigma^2 u \bar{u}+e^2 \bar{u}^2 \left(D_\mu u \right)^2 \sigma^2+\right.\nonumber\\
&&\left.+e^2 u^2 \left(D_\mu \bar{u} \right)^2 \sigma^2\right)+\text{(fermions)},\\
\mathcal{L}_{CS}&=&\frac{\kappa}{4\pi}\left(\epsilon^{\mu\nu\rho}A_\mu \pa_\nu A_\rho+2 D\sigma\right)+\text{(fermions)}.
\eea

Note that again the extra field $\sigma$, necessary for the extended SUSY, is non-dynamical and therefore its equation of motion is purely algebraic. As a consequence, there is an extra constraint in the set of auxiliary field equations 
\be
\frac{\pa \mathcal{L}}{\pa D}=0,\,\,\,\frac{\pa \mathcal{L}}{\pa F}=0 \,\,\,\text{and}\,\,\, \frac{\pa \mathcal{L}}{\pa \sigma}=0\label{sys}.
\ee

The system (\ref{sys}) possesses two solutions. One of them (which we call the trivial phase) involves a trivial solution for $F$ ($F=0$) and leads to higher-order time derivatives (see Appendix B). In the second solution (the nontrivial phase) the auxiliary fields take the form (without fermions),
\bea
D &=&\frac{8\pi^2}{\kappa^2}e^3  h\Omega \Delta \label{aux1}, \\
\sigma &=&-\frac{2\pi}{\kappa}e\Omega \label{aux2},\\
F\bar{F} &=& -\frac{g(u\bar{u})}{2 h(u\bar{u})}-D_\mu u D^\mu \bar{u}+4\pi^2  \left(\frac{e^2}{\kappa}\right)^2 \vert u \vert^2\Omega^2 \label{aux3},
\eea
where
\be
\Omega=\frac{1}{2}\left(u \pa_ u\mathcal{K}+\bar{u}\pa_{\bar{u}}\mathcal{K}\right),\quad \Delta=\pa_\mu \left(u\bar{u}\right) \pa^\mu\left(u\bar{u}\right),\quad g=\pa^2_{u\bar{u}}\mathcal{K}.\label{defk}
\ee

From now on, we assume that all the equations refer only to the bosonic sector. After using (\ref{aux1})-(\ref{aux3}), we arrive at the on-shell Lagrangian 
\bea
\mathcal{L}=&&\frac{\kappa}{4\pi}\epsilon^{\mu\nu\rho}A_\mu\pa_\nu A_\rho+h\left(\left(D_\mu u\right)^2\left(\overline{D_\nu u}\right)^2-\left(D_\mu u \overline{D_\mu u}\right)^2\right)\label{CSbS}\\
&-&\frac{g^2}{4h}+\frac{4e^4\pi^2}{\kappa^2}h\Omega^2\Delta.\nonumber
\eea
We will refer to (\ref{CSbS}) as the modified Chern-Simons baby Skyrme model. The first line in (\ref{CSbS}) corresponds to the Chern-Simons term and the BPS baby Skyrme term. The first term of the second line is the ``emergent" potential, which, as it happens in the usual $N=2$ BPS baby Skyrme model, appears without the need of including F-terms in the action. This potential is constructed in terms of the target space area density $h$ and the K\"{a}hler metric $g$ of the $\sigma$-model part. In addition, the $\sigma$-model term (first term in (\ref{bSCS1})) is absent from the on-shell action. Because of this, changes in the K\"ahler metric only have an impact in the potential. This phenomenon is not new and also happens in the gauged baby Skyrme model with Maxwell term \cite{Queiruga2}. The relevant fact here is that as a consequence of the extended supersymmetry a new part has been added to the Lagrangian, namely the last term in (\ref{CSbS}). This term is quadratic in derivatives and, as we will see later, it is essential to ensure the existence of self-dual equations. 

In general situations it is difficult to provide a recipe to bring an action from non-BPSness to BPSness. That is to say, what kind of terms we have to add in order to have solutions satisfying Bogomoln'yi equations? On the other hand, the construction of a SUSY extended version of a model, when possible, implies BPSness of the underlying bosonic model and, at the same time it provides the BPS equations. As we will show below, the presence of the last term in (\ref{CSbS}) (containing two derivatives) is neccessary to have BPS structure and it is merely a consequence of the auxiliary field equations (\ref{sys}) imposed by the $\mathcal{N}=2$ SUSY. It should be noted that (\ref{bSCS1}) can be supersymmetrized in the $\mathcal{N}=1$ superspace without any modification, but the resulting model is not BPS.


\subsection{Energy bound and BPS equations}
\label{BPS}

Since in Chern-Simons systems the solutions carry electric and magnetic charge, we cannot take the solution $A_0=0$ for the time component of the gauge field. In this case, $A_0$ can be solved via its equation of motion and, in the static case has the form
\be
A_0=-\frac{\kappa}{4e^2\pi}\frac{F_{12}}{h\Delta_s},
\ee
where $\Delta_s=\pa_i \left( u \bar{u}\right)\pa_i \left( u \bar{u}\right)$. The static Lagrangian can be written as
\bea
\mathcal{L}\vert_{\text{static}}=&-&\frac{\kappa^2}{16 e^2 \pi^2}\frac{F_{12}^2}{h\Delta}+h\left(\left(D_i u\right)^2\left(\overline{D_j u}\right)^2-\left(D_i u \overline{D_i u}\right)^2\right)\label{staL}\\
&-&\frac{g^2}{4h}-\frac{4e^4\pi^2}{\kappa^2}h\Omega^2\Delta\nonumber.
\eea

After some algebra we can write the energy functional as follows
\bea
E&=&\int d^2x\left( h\left(i \mathcal{T} \pm \frac{g}{2h}\right)^2+\frac{\kappa^2}{16 e^2\pi^2h\Delta}\left(F_{12}\pm\frac{8\pi^2}{\kappa^2}e^3 h\Omega\Delta\right)^2\right)\label{enfun}\\
&\mp&\int d^2x\left(i Q g+e \Omega F_{12}\right).\nonumber
\eea

Since the first line in (\ref{enfun}) is positive the energy has a lower bound
\be
E\geq \vert \int d^2x\left(i \mathcal{T} g+e \Omega F_{12}\right) \vert.\label{Ebound}
\ee

This bound is saturated when both terms in brackets in the first line of (\ref{enfun}) vanish. It is easy to verify that the self-dual equations (see appendix C)
\bea
i\mathcal{T}&=&\pm \frac{g}{2 h}\label{Bogo1},\\
 F_{12}&=&\pm\frac{8\pi^2}{\kappa^2}e^3 h\Omega\Delta \label{Bogo2},
\eea
also satisfy the Euler-Lagrange equations. 
Of course the procedure to obtain the self-dual equations (\ref{Bogo1}) and (\ref{Bogo2}) can be a difficult task in general.  But, as the model has extended supersymmetry we can apply our general BPS equations (\ref{gen1})-(\ref{sigma0}). By substituting (\ref{aux1})-(\ref{aux3}) in (\ref{gen1}) and (\ref{gen2}) we obtain (\ref{Bogo1}) and (\ref{Bogo2}). Hence, once an off-shell $\mathcal{N}=2$ formulation of the model is known, the computation of the BPS equations is again simple. Moreover, taking into account that $F\neq0$  (\ref{aux3}), we obtain from Table \ref{BPStype} that the solutions of (\ref{Bogo1}) and (\ref{Bogo2}) are $1/4$-BPS.




\section{Maxwell Chern-Simons baby Skyrme}
\label{YMCSbS}

The $\mathcal{N}=2$ Maxwell term in the $\mathcal{N}=2$ superspace has the following form
\be
\mathcal{L}_{YM}=\int d^2\theta W^\alpha W_\alpha +\text{h.c.}=-\frac{1}{4}F_{\mu\nu}F^{\mu\nu}+i\bar{\lambda}^\alpha \pa_\alpha^{\,\beta}\lambda_\beta+\frac{1}{2}D^2+\frac{1}{2}\pa^\mu\sigma\pa_\mu\sigma.\label{YM}
\ee

The coupling of (\ref{YM}) to the gauge invariant baby Skyrme model does not spoil the BPS structure \cite{Queiruga2}. In this model, the  field $\sigma$ is dynamical but the vacuum solution $\sigma=0$ solves the field equation. We now consider the Lagrangian
\be
\mathcal{L}=\mathcal{L}_\sigma+\mathcal{L}_4+\mathcal{L}_{CS}+\mathcal{L}_{M}.
\ee

Also in this case, due to the presence of the Yang-Mills term the field $\sigma$ is dynamical, and due to the Chern-Simons term, the ``vacuum" solution ($\sigma=0$) does not solve the Euler-Lagrange equations. The first conclusion we can draw from this fact is that the combination of the $\mathcal{N}=2$ Yang-Mills and Chern-Simons terms requires an extra nontrivial dynamical degree of freedom $\sigma$. But since the model is still $\mathcal{N}=2$, the BPS structure is preserved. The auxiliary field equations give
\bea
D&=&-e\Omega-\frac{\kappa \sigma}{2\pi}\label{BPS1},\\
F\bar{F}&=&-\frac{g}{2h}-D^\mu\bar{u}D_\mu u +e^2\vert u \vert^2\sigma^2.\label{BPS2}
\eea
Once (\ref{BPS1}) and (\ref{BPS2}) are taken into account the component action reads
\bea
\mathcal{L}=&&\frac{\kappa}{4\pi}\epsilon^{\mu\nu\rho}A_\mu\pa_\nu A_\rho-\frac{1}{4}F_{\mu\nu}F^{\mu\nu}+h\left(\left(D_\mu u\right)^2\left(\overline{D_\nu u}\right)^2-\left(D_\mu u \overline{D_\mu u}\right)^2\right)\label{CSbSMM}\\
&&+\frac{1}{2}\pa_\mu\sigma\pa^\mu\sigma+e^2h\sigma^2\Delta-V(u,\bar{u},\sigma),\nonumber
\eea
where
\be
V(u,\bar{u},\sigma)=\frac{g^2}{4h}+\frac{e^2}{2}\Omega^2+\frac{e\kappa}{2\pi}\sigma\Omega+\frac{\kappa^2}{8\pi^2}\sigma^2.
\ee
As we have announced, the neutral field action cannot be eliminated from the action and it is now a genuine dynamical field. The auxiliary fields (\ref{BPS1}) and (\ref{BPS2}) in combination with (\ref{gen1})-(\ref{gen2}) provide the BPS equations of the model
\bea
F_{12}&=&\pm \left(e\Omega+\frac{\kappa \sigma}{2\pi}\right),\label{BPS11}\\
iQ&=&\pm \frac{g}{2 h}.\label{BPS22}
\eea

  Interestingly enough, the matter BPS equation (\ref{BPS22}) does not change, while the gauge BPS equation (\ref{BPS11}) involves now the extra neutral field $\sigma$. In the summary we briefly discuss the possible limits of this model. As in the previous case, due to (\ref{BPS2}) the solutions of (\ref{BPS11}) and (\ref{BPS22}) are $1/4$-BPS. Moreover, there is also a trivial phase ($F=0$) with higher-time derivatives and 1/2-BPS states (see appendix B). In table \ref{classification} we summarized the classification of the models discussed here.

\begin{table}
\begin{center}
  \begin{tabular}{ | c | c | c | c | }
    \hline
   model & $\mathcal{N}=1$ & $\mathcal{N}=2$&BPS-type \\ \hline
   GNL$\sigma$CS+BPS potential &Yes&Yes&$1/2$\\ \hline
   GNL$\sigma$CS+arbitrary potential&Yes&No&No\\ \hline
     $\mathbb{C}P^1$-type+CS model (\ref{zak})  &Yes&No&No\\ \hline
    GbSCS&Yes&No&No\\ \hline
    Modified GbSCS (nontrivial phase)&Yes&Yes&$1/4$\\ \hline
    Modified GbSCSM (nontrivial phase)&Yes&Yes&$1/4$\\ \hline
       Modified GbSCS (trivial phase)    &Yes&Yes&$1/2$\\ \hline
       Modified GbSCSM (trivial phase)    &Yes&Yes&$1/2$\\ \hline

  \end{tabular}
  \caption{Classification of the models discussed in the paper according to SUSY and BPS-type. GNL$\sigma$CS: gauged nonlinear $\sigma$-model with Chern-Simons term. Modified GbSCS: Modified gauged baby Skyrme model with Chern-Simons term. Modified GbSCSM: Modified gauged baby Skyrme model with Chern-Simons and Maxwell terms.}
    \label{classification}  
\end{center}

\end{table}



\newpage
\section{Summary}

In this work we have studied the SUSY version of various models containing the Chern-Simons term. We have shown that the coupling of the Chern-Simons term to a general nonlinear $\sigma$-model allows for general potentials (depending on the K{\"a}hler metric) and preserves two supersymmetries. We have also constructed an ``universal" form for the set of BPS equations for supersymmetric gauge models in three dimensions. This set of equations is obtained directly from the supersymmetric variation of the fermions and depends only on the complex auxiliary fields $F_a$ (from the chiral  superfields) and the real auxiliary fields $D_a$ (from the vector superfields). These equations are verified by all BPS solutions, no matter the amount of SUSY they preserved. Moreover, within the same framework, we are able to determine the BPS-type (number of unbroken SUSY generators) of the solution, simply by computing the kernel of the on-shell fermionic transformations. 

Then, we focussed on the SUSY gauged Chern-Simons $\mathbb{CP}^N$ models and the SUSY form of the so-called ``geometric formulation" of the $\mathbb{CP}^N$ model was reviewed. This is simply a particular case of the general SUSY form of the gauge nonlinear $\sigma$-models \cite{Witten2}. The SUSY form of the so-called ``gauged formulation" is more subtle. Once the dynamical gauge fields are added, the model presents an $U(1)\times U(1)$ local symmetry. One part comes from the ``standard" gauge symmetry (the one associated with the dynamical gauge fields) and the other is necessary to reduce one d.o.f. in the set of complex fields. In the superspace formulation this results in the introduction of two vector supefields (\ref{cpngauge2}), one produces the kinetic term for the gauge fields and the other implements the constraint on the complex fields. We have provided also various applications of the SUSY form: direct computation of the self-dual potential and self-dual equations, a comparison with previous non-supersymmetric results and determination of the BPS-type. The same procedure, including computation of BPS potentials, BPS bounds and type can be straightforwardly applied to the so-called ``generalized models," see for example \cite{Bazeia1,Bazeia2,Bazeia3}.

In the second part of the paper we studied a particular example of SUSY higher-derivative field theory, the baby Skyrme model gauged by a Chern-Simons term. Despite the fact that the BPS baby Skyrme model (even in its gauged version coupled to a Maxwell term) possesses a BPS structure, the coupling to a Chern-Simons term in an $\mathcal{N}=2$ invariant way requires an extra quadratic term in derivatives. This term is dictated by the extended SUSY and its absence reduces the supersymmetry to $\mathcal{N}=1$ and breaks the BPS structure. 


We have also studied the Maxwell Chern-Simons baby Skyrme model. The addition of a Maxwell term makes the neutral field $\sigma$ dynamical. Therefore, it turns out that, in order to have $\mathcal{N}=2$ (and as a consequence, BPS equations) a formulation in terms of the complex field $u$ alone is impossible. Regarding the limiting models we have different situations. In the limit $\kappa\rightarrow \infty$, we can neglect the kinetic term for the field $\sigma$ and consider it again as an auxiliary field just reproducing (\ref{CSbS}). In the limit $\kappa\rightarrow 0$, the Chern-Simons term is absent and we can take again $\sigma=A_0=0$, leading the the Maxwell baby-Skyrme. It is also interesting to note that for the solution (\ref{BPS2}), the model is not connected with the standard Maxwell Chern-Simons Higgs, since the limit $h\rightarrow 0$, causes a divergence in the auxiliary field $F$. In the nontrivial phase, $F\neq0$, the BPS solutions of both models (with and without Maxwell term) only preserve $1/4$ of SUSY. The trivial phase ($F=0$) gives the standard Maxwell Chern-Simons Higgs in the same limit, but for $h\neq 0$ the model contains higher-order time derivatives (see Appendix B). In both cases, the BPS solitons are of $1/2$-type. This result can be extended to general higher-derivative corrections. As far as these terms are of at least quadratic order in $F$, the $F=0$ solution is always available and the BPS solitons will be $1/2$-BPS.

It is interesting to note that both the baby Skyrme model, and the gauged baby Skyrme model (with a Maxwell term) preserve the SDiff symmetry. It is widely believe that this symmetry is closely related to the BPS structure (as it is the case for these two models). On the other hand, the existence of the BPS equations for the Chern-Simons baby Skyrme model requires the presence of a term, $\Delta$, which breaks explicitly the SDiff symmetry while preserving the conformal symmetry. These both symmetries are the most common symmetries for solitons in $2+1$ dimensions. For the modified Chern-Simons baby Skyrme model, it is precisely the introduction of a SDiff breaking term, what brings the model back to the BPS limit. We also recall that the $\Delta$ term can be rewritten as a generalized kinetic term for $\phi_3$ (the third component of the unit vector in the $O(3)$ formulation) of the form $f(\phi_3)\pa_\mu\phi_3\pa^\mu\phi_3$.

Narrowing down the relation between extended SUSY and BPS, it would be also interesting to analyze what happens when the geometry of the base manifold is modified. It is well-known that for some higher-derivative models without BPS solutions in $\mathbb{R}^3$ the change of geometry allows for the existence of BPS solitons for a finite number of topological charges \cite{manton,adamw}. This should be translated into the existence of extended SUSY only in some topological sectors. This issue is under current investigation. We finally remark that, $\mathcal{N}=2$ in three dimensions is not only an useful tool in the study of BPS solutions but also can be used as a guiding principle to build new BPS models.


\vspace{0.5cm}

{\bf Acknowledgements.}- It is a pleasure to thank A. Wereszczynski and C. Adam for useful comments in a previous version of the manuscript.



\appendix

\section{$\sigma$ equation}

Here we show that the equation (\ref{sigma0})  is a consequence of equation (\ref{gen2}) and the $\mathcal{N}=2$ SUSY. The dimensional reduction from $3+1$ dimensions provides an useful correspondence between $\mathcal{N}=1$, $d=3+1$ theories and $\mathcal{N}=2$, $d=2+1$ theories. In this reduction, the third component of the gauge field is transformed into the neutral field $\sigma$, 
\be
A_{\mu=0,1,2,3}\rightarrow A_{\mu=0,1,2}\oplus\sigma,\quad \pa_3 G=0,
\ee
where $G$ is any function depending on the fields. This has the consequence that as far as we consider reduced terms coming from $3+1$ dimensions, the static Euler-Lagrange equations for $A_0$ and $\sigma$ are functionally the same up to a sign. If $EL_X$ denotes the Euler-Lagrange operator acting on the field $X$ and $\mathcal{L}^{\text{red}}$ the part of the Lagrangian coming form the dimensional reduction we have
\be
EL_{A_0}[\mathcal{L}^{\text{red}}]=-EL_{\sigma}[\mathcal{L}^{\text{red}}]\vert_{\sigma\rightarrow A_0}.\label{sr1}
\ee
Moreover, for the actions consider here, the field equations for $A_0$ and $\sigma$ are linear on these fields and therefore,
\be
EL_{A_0}[\mathcal{L}^{\text{red}}]=-EL_{-A_0}[\mathcal{L}^{\text{red}}],\quad EL_{\sigma}[\mathcal{L}^{\text{red}}]=-EL_{-\sigma}[\mathcal{L}^{\text{red}}].\label{sr2}
\ee
Regarding the Chern-Simons term the situation is different. It does not come from any dimensional reduction form four dimensions, and $A_0$ and $\sigma$ play different roles, specifically
\be
EL_{A_0}[\mathcal{L}_{CS}]=\frac{\kappa}{2\pi}F_{12},\quad EL_{\sigma}[\mathcal{L}_{CS}]=\frac{\kappa}{2\pi}D\label{sr3}.
\ee
Now taking into account (\ref{sr1})-(\ref{sr3}) it is easy to see that both fields satisfy the same equation up to a sign provided that (\ref{gen2}) is verified.


\section{The trivial phase}

As we announced in Sec. \ref{SUSY}, the auxiliary field equations for the model (\ref{bSCS2}) have more than one solution due to the nonlinearity of the equation on $F$. The solutions presented in Sec.  \ref{SUSY}, correspond to the {``}nontrivial" phase ($F\neq 0$). These solutions are more interesting for two reasons: first, they lead to an action containing the baby-Skyrme term, and second, they do not produce higher-order time derivatives. For completeness, we give here explicitly the solution for the {``}trivial" phase:
\bea
D &=&\frac{8e^3\pi^2\vert u\vert^2\Omega}{\kappa^4}\left(8e^4 \pi^2h \vert u\vert^2 \Omega^2-g\kappa^2\right) \label{aux11}\nonumber \\
&+&\frac{8e^3\pi^2h\Omega}{\kappa^2}\left(\bar{u}^2\left(D_\mu u\right)^2+u^2\left(D_\mu \bar{u}\right)^2\right),\\
\sigma &=&-\frac{2\pi}{\kappa}e\Omega ,\label{aux21}\\
F &=& 0, \label{aux31}
\eea
and the resulting Lagrangian
\bea
\mathcal{L}\vert_{\text{t.p.}}&=&g (D_\mu u D^\mu \bar{u})+h \left(D_\mu u\right)^2\left(D_\nu \bar{u}\right)^2+\frac{4 e^4 \pi^2 h\Omega^2}{\kappa^2}\left(\bar{u}^2\left(D_\mu u\right)^2+u^2\left(D_\mu \bar{u}\right)^2\right)\nonumber\\
&+&\frac{4e^4\pi^2\vert u\vert^2\Omega^2}{\kappa^4}\left(4e^4 \pi^2h \vert u\vert^2 \Omega^2-g\kappa^2\right)+\frac{\kappa}{4\pi}\epsilon^{\mu\nu\rho}A_\mu F_{\nu\rho}.\label{tplag}
\eea
The first obvious difference with (\ref{CSbS}) is that the $\sigma$-model term is still present in the on-shell Lagrangian. Also, as we announced before, this new Lagrangian (\ref{tplag}) does not contain the baby Skyrme term, but instead, a quartic term in time derivatives (the second term in the first line of (\ref{tplag})). Despite all this, the model still possesses a BPS structure. The BPS equations are
\be
D_i u=\pm i D_2 u,\quad F_{12}=\pm \frac{8 e^3 \pi^2\Omega}{\kappa^2}\left(g\vert u \vert^2+\bar{u}^2\left(D_i u\right)^2+u^2\left(D_i\bar{u}\right)^2\right).\label{appBPS}
\ee

Moreover, since $F=0$ the solutions of (\ref{appBPS}) are of $1/2$-type.

\section{Euler-Lagrange and BPS equations}

The static Euler-Lagrange equations for the Lagrangian (\ref{CSbS}) are
\be
\frac{\kappa^2}{8 e^2 \pi^2}\pa_i\left(\frac{F_{12}}{h \Delta}\right)=2 i e h \pa_i(u\bar{u}) \mathcal{T},\label{EL1}
\ee
for the gauge field and
\bea
&&2\epsilon_{ij}\bar{D}_i \left(\mathcal{T} h D_j \bar{u}\right)-\pa_u h \mathcal{T}^2-\frac{\kappa^2}{16 e^2 \pi^2} \left( \frac{F_{12}^2}{h\Delta^2} 2\bar{u}\pa_j^2 (u\bar{u})- \frac{F_{12}^2}{h^2\Delta}\pa_u h\right)-\nonumber\\
&&-\frac{4 e^4\pi^2}{\kappa^2}\left(\pa_u h \Omega^2 \Delta+2 h \Omega \pa_u\Omega \Delta-2 h \Omega^2 \bar{u}\pa_j^2(u\bar{u})\right)+ \nonumber\\
&&+\frac{g\pa_u g}{2 h}-\frac{g^2}{4 h^2}\pa_u h=0,\label{EL2}
\eea
for the complex scalar field. It is easy to verify that the BPS equations (\ref{Bogo1}) and (\ref{Bogo2}) imply (\ref{EL1}). From the definition of $\Omega$ and $g$ (\ref{defk}) we have
\be
\pa_i\Omega=g \pa_i(u\bar{u}).\label{gome}
\ee
Therefore, after differentiating both sides in (\ref{Bogo2}) and using (\ref{gome}) and (\ref{Bogo1}) we arrive at (\ref{EL1}). In order to verify that the BPS equations imply  (\ref{EL2}) let us define
\bea
X_1&=&-\frac{\kappa^2}{16 e^2 \pi^2}\frac{F_{12}^2}{h\Delta^2} 2\bar{u}\pa_j^2 (u\bar{u})+\frac{8 e^4\pi^2}{\kappa^2} h \Omega^2 \bar{u}\pa_j^2(u\bar{u}),\\ 
X_2&=&-\pa_u h \mathcal{T}^2-\frac{g^2}{4 h^2}\pa_u h,\\ 
X_3&=&\frac{\kappa^2}{16 e^2 \pi^2} \frac{F_{12}^2}{h^2\Delta}\pa_u h-\frac{4 e^4\pi^2}{\kappa^2}\pa_u h \Omega^2 \Delta,\\ 
X_4&=&-2ie \mathcal{T}hF_{12}\bar{u}-\frac{8 e^4\pi^2}{\kappa^2} h \Omega \pa_u\Omega \Delta,\\
X_5&=&\epsilon_{ij}\pa_i \left(\mathcal{T} h\right) D_j \bar{u}+\frac{g\pa_u g}{2 h}.
\eea
In terms of $X_i$, (\ref{EL2}) can be written as $X_1+X_2+X_3+X_4+X_5=0$. Direct substitution of (\ref{Bogo2}) gives $X_1=0$ and $X_3=0$, and direct substitution of (\ref{Bogo1}) gives $X_2=0$. $X_4=0$ follows from the substitution of (\ref{Bogo1}) and (\ref{Bogo2}) and the relation $\pa_u\Omega=g\bar{u}$. Finally $X_5=0$ follows from two substitutions of (\ref{Bogo1}).

\end{document}